\documentclass[10pt,twocolumn]{article}

\usepackage{multicol}
\usepackage{subfigure}
\usepackage{graphicx}
\usepackage{amsmath}
\usepackage{lipsum}
\usepackage{color}
\usepackage{siunitx}
\usepackage{authblk}
\usepackage[margin=2cm]{geometry}
\usepackage[numbers, square, super]{natbib}

\title{Wetting morphologies on an array of fibers of different radii}

\author[1]{Alban Sauret}
\author[2]{Fran\c{c}ois Boulogne}
\author[3]{David C\'ebron}
\author[4]{Emilie Dressaire}
\author[2]{Howard A. Stone}
\affil[1]{  ~Surface du Verre et Interfaces, UMR 125, 93303 Aubervilliers, France. E-mail: alban.sauret@saint-gobain.com.}
\affil[2]{ ~Department of Mechanical and Aerospace Engineering, Princeton University, Princeton, New Jersey 08544, USA. E-mail: hastone@princeton.edu.}
\affil[3]{ ~Universit\'e Grenoble Alpes, CNRS, ISTerre, Grenoble, France.}
\affil[4]{ ~Department of Mechanical and Aerospace Engineering, New York University Polytechnic School of Engineering, Brooklyn, NY 11201, USA. }

\date{\today}
\begin{document}


\twocolumn[
    \begin{@twocolumnfalse}
        \maketitle
        \begin{abstract}
            We investigate the equilibrium morphology of a finite volume of liquid placed on two parallel rigid fibers of different radii. As observed for identical radii fibers, the liquid is either in a column morphology or adopts a drop shape depending on the inter-fiber distance. However the cross-sectional area and the critical inter-fiber distance at which the transition occurs are both modified by the polydispersity of the fibers. Using energy considerations, we analytically predict the critical inter-fiber distance corresponding to the transition between the column and the drop morphologies occurs. This distance depends both on the radii of the fibers and on the contact angle of the liquid. We perform experiments using a perfectly wetting liquid on two parallel nylon fibers: the results are in good agreement with our analytical model. The morphology of the capillary bridges between fibers of different radii is relevant to the modeling of large arrays of polydisperse fibers.
        \end{abstract}
    \end{@twocolumnfalse}
]

\section{Introduction}

The wetting of rigid and elastic fibers is present in many natural and engineered systems: paper\cite{Hubbe2006} and textiles,\cite{Minor1959,Kissa1996,Monaenkova2013} filters\cite{Contal2004,Sutter2010}, fog-harvesting nets,\cite{Park2013} human and animal hair\cite{Rijke1968,Dawson1999,Eisner2000,Rijke2010,Zheng2010,Srinivasan2014} or, more recently, microfabricated systems.\cite{Chandra2009,Pokroy2009} Indeed, various biological or technological materials are modeled as a network of fibers that are parallel or randomly oriented. The presence of ambient humidity,\cite{Zheng2010,Ju2012} the impact of a drop\cite{Lorenceau2004,Piroird2009} or the addition of a fluid leads to the formation of capillary bridges similar to to those observed for liquids in granular materials.\cite{Herminghaus2005,Kudrolli2008,Strauch2012} The ability to model networks of fibers partially saturated with a liquid is crucial to determine their mechanical properties. A key point for such modeling is the ability to describe the liquid morphology between fibers.

Fundamental studies have considered the wetting\cite{Quere1988,Quere1999} and the equilibrium shape of a drop lying on a single fiber.\cite{carroll1976,carroll1986,McHale2001} Based on energetic considerations, it has been shown that a drop of liquid deposited on a fiber can either be in an axisymmetric barrel morphology engulfing the fiber or an asymmetric clamshell morphology with the droplet sitting on the side of the fiber.\cite{Chou2011} Methods involving electrowetting effects have also been considered to tune the shape of the drop.\cite{Eral2011,DeRuiter2012}

However, understanding the wetting of complex fibrous media first requires modeling the morphology of capillary bridges between several fibers. Princen has characterized the column morphology that the liquid adopts on two or more parallel fibers of the same radii.\cite{Princen1970} In the column morphology, the liquid is confined between the two fibers and has a constant cross-section all along the column. If some liquid is added or removed from the column, it will grow or shrink in length while its cross-section is not modified. Princen provided an analytical description of the shape of the liquid cross section in the column morphology, which makes amenable a prediction of the existence of a liquid column at small inter-fiber distances. Above a critical distance between the fibers, the liquid collects in a drop, which overspills the two fibers, as can be observed experimentally.\cite{Miller1967,Protiere2012} The experiments with a drop on two rigid parallel fibers have shown that the transition between the drop and column morphologies is hysteretic. More recently, the equilibrium morphologies of the liquid on crossed fibers have also been studied both analytically and experimentally.\cite{Sauret2014} In this situation, a third morphology is possible: the mixed morphology where a drop on one side together with a column of liquid on the other side coexist. Whereas these different studies considered one or two rigid fibers, the competition between the elasticity of the fibers and the capillary effects has also been investigated,\cite{Duprat2012} showing that a bundle of fibers organizes into clusters.\cite{Bico2004,Py2007,Liu2012}

To the best of our knowledge, all of these studies devoted to modeling the wetting of fibers focused on a pair or a large assembly of identical fibers. Yet, many biological or industrial systems, such as glass wool, involve a network of polydisperse fibers, i.e., fibers with different radii. In previous work, the capillary bridge between two or more spheres of different radii has been described, as well as its influence on the volume or the force exerted between the spheres\cite{Soulie2006,Scheel2008,Delenne2015} but the particular situation of fibrous material remains to be considered. Indeed, the difference in fiber size is expected to affect the morphology and stability of the capillary bridge.

In this paper, we build on the seminal study conducted by Princen\cite{Princen1970} and we consider the morphology of a small volume of liquid placed on two parallel fibers of different radii, $a_1 > a_2$. In principle, this study could be performed for fibers randomly oriented but the calculation would be more cumbersome without adding to the physical picture. As a result, we focus on two parallel fibers with different radii and present an analytical description of the shape of the liquid cross-section in the column morphology. We also determine the critical inter-fiber distance at which the column to drop transition occurs. The results of the analytical model are then compared with experimental data obtained with a perfectly wetting fluid. The good agreement between the predictions and experiments validates our approach and our analytical modeling.


\section{Analytical modeling} \label{sec:Analytical_Modeling}

\begin{figure}
    \centering
 \subfigure[]{\includegraphics[width=8cm]{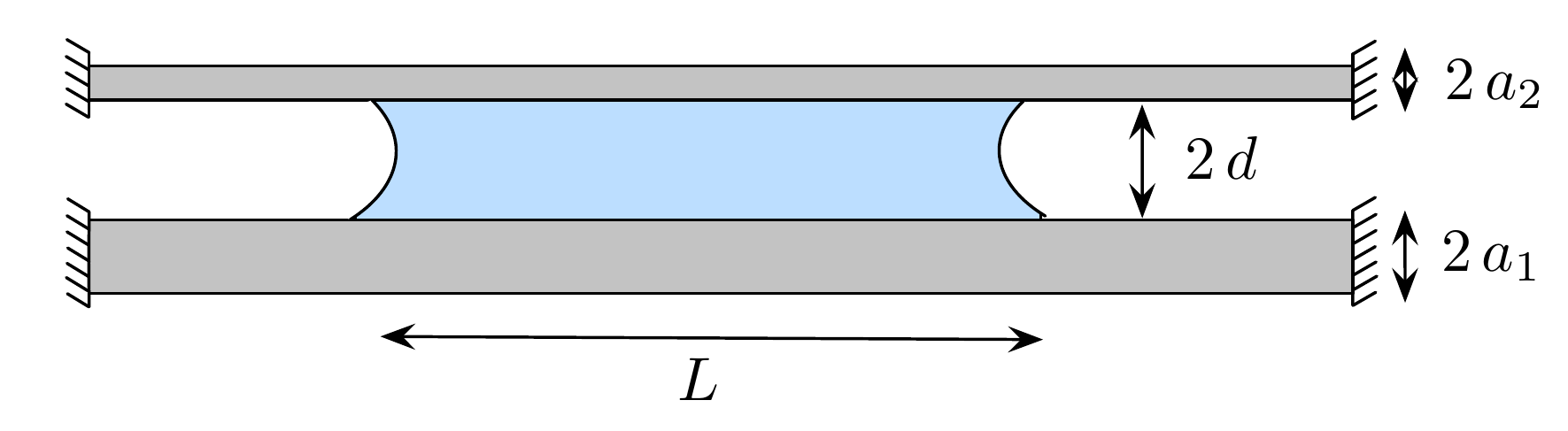}}
  \subfigure[]{\includegraphics[width=7cm]{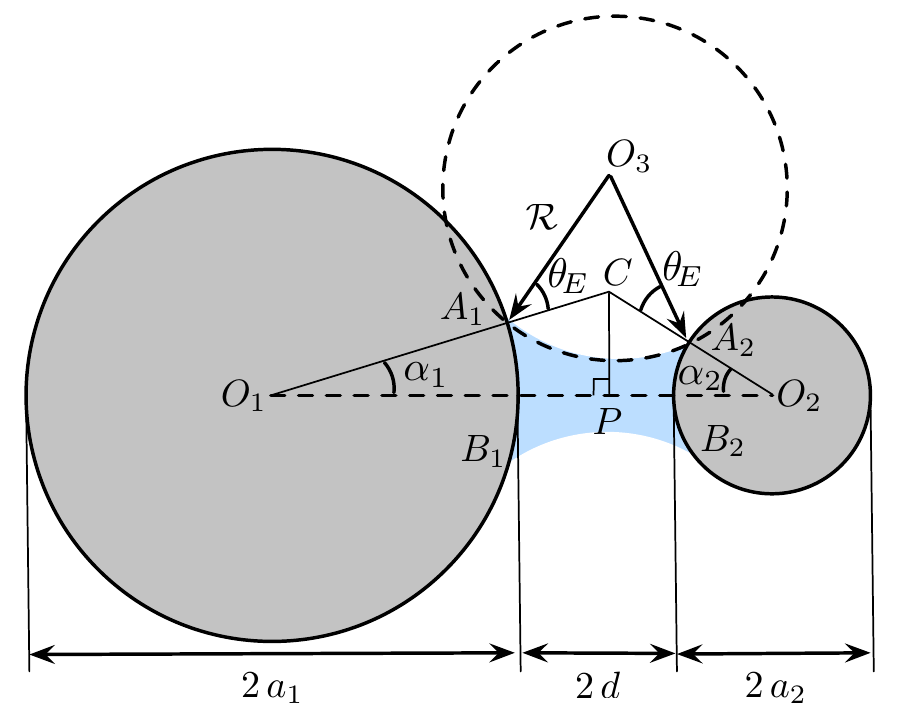}}
\caption{(a) Top view schematic of the system composed of two fibers of radii $a_1$ and $a_2$ separated by a distance $2 d$. (b) Cross-section of a liquid column where $\alpha_1$ and $\alpha_2$ denote the wrapping angles of the fluid around the fibers and $\theta_E$ is the liquid-fiber contact angle.}\label{fig:schema}
\end{figure}

We consider two rigid fibers of radii $a_1$ and $a_2$ ($a_1>a_2$) and constant cross-section whose respective axes are denoted ($O_1\,z$) and ($O_2\,z$). The fibers are parallel and separated by a distance $2d$ (see figure \ref{fig:schema}(a)). If the distance between the fibers is smaller than a critical distance $2d_c$, we assume that a wetting liquid, with contact angle $\theta_E<90^{\rm o}$, deposited on the fibers spreads into a column shape. As shown by Princen,\cite{Princen1970} the column has a constant cross-section except at the two terminal menisci whose analytical description remains difficult. However, for a sufficiently long column, assuming a constant cross-section provides a good approximation. The perpendicular cross-section of a liquid bridge between two fibers of radii $a_1$ and $a_2$ ($a_1>a_2$) is schematically represented in figure \ref{fig:schema}(b). The shape of the cross-section is characterized by its radius of curvature $\mathcal{R}$ and the wrapping angles of the fluid around the large and small fibers, $\alpha_1$ and $\alpha_2$ respectively.

\subsection{Geometrical determination}

Because two inscribed triangles share a common side $CP$ (figure \ref{fig:schema}(b)), the relation between the wrapping angles $\alpha_1$ and $\alpha_2$ is described by
\begin{equation}
\frac{\sin \alpha_1}{\sin \alpha_2}=\frac{O_2\,C}{O_1\,C}=\frac{a_2+CA_2}{a_1+CA_1} \mbox{.}
\end{equation}
In addition, the distance between the axes of the two fibers $O_1\,O_2$ is given by:
\begin{eqnarray}
O_1\,O_2 & = & a_1+a_2 +2d = P\,O_1+P\,O_2 \nonumber  \\
& = & (a_1+CA_1)\cos\alpha_1+(a_2+CA_2)\cos\alpha_2   \mbox{,} \nonumber \\
\end{eqnarray}
which leads to a relation between $\alpha_1$, $\alpha_2$ and $d$:
\begin{equation}
C\!A_1\cos\alpha_1+C\!A_2\cos\alpha_2 \!=\! a_1(1-\cos\alpha_1)+a_2(1-\cos\alpha_2)+2 d  \mbox{.}
\end{equation}
The triangle $A_1O_3A_2$ is isosceles with $\mathcal{R}=O_3A_1=O_3A_2$. In addition, $O_3C$ defines the bisector of angle $A_1O_3A_2$. As a result, $CA_1$ and $CA_2$ are equal:
\begin{equation}
CA_1=CA_2=\frac{a_1(1-\cos\alpha_1)+a_2(1-\cos \alpha_2)+2d}{\cos\alpha_1+\cos\alpha_2} \mbox{.}
\end{equation}

\begin{figure}
    \centering
\includegraphics[width=3.5cm]{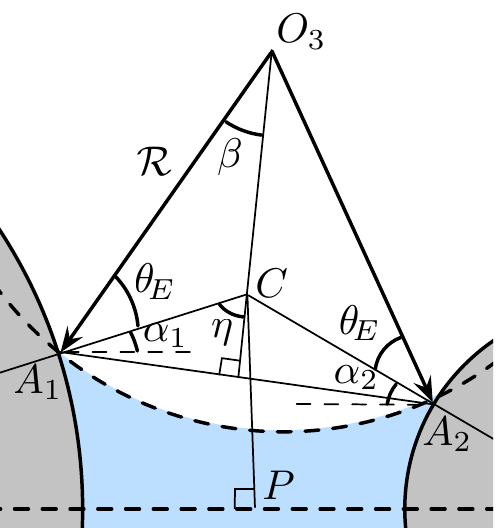}
\caption{Focus on the cross-section of a liquid column.}\label{fig:zoom_schema}
\end{figure}

The radius of curvature $\mathcal{R}$ of the cross-section is given by
\begin{equation}
\mathcal{R}  =  O_3A_1=CA_1\,\frac{\sin \eta}{\sin \beta}  \mbox{,} \label{has1}
\end{equation}
where $\eta$ and $\beta$ are two angles defined in figure \ref{fig:zoom_schema}. Geometrically, we find that $\eta=(\pi-\alpha_1-\alpha_2)/2$ and $\beta=(\pi-\alpha_1-\alpha_2-2\,\theta_E)/2$.
Using the radius of the large fiber, $a_1$, as the characteristic length scale and the ratio of the fiber radii, denoted $\Gamma=a_2/a_1$, equation (\ref{has1}) can be written:
\begin{equation}
\mathcal{\tilde{R}}=\frac{\mathcal{R}}{a_1} =\frac{(1-\cos\alpha_1)+\Gamma\,(1-\cos \alpha_2)+2\,\tilde{d}}{\cos(\alpha_1+\theta_E)+\cos(\alpha_2+\theta_E)} \mbox{,} \label{has22}
\end{equation}
where $\tilde{d}=d/a_1$. We also need to determine the cross-sectional area $\mathcal{A}$ of the liquid column. A calculation based on the geometry of the system is detailed in the appendix and leads to
\begin{eqnarray}
\frac{\mathcal{A}}{{a_1}^2}  = \mathcal{\tilde{R}}[{\cos(\alpha_1+\theta_E)+\cos(\alpha_2+\theta_E)}]\left(\sin\alpha_1+\Gamma\,\sin\alpha_2\right) \nonumber \\
-\left(\alpha_1-\sin \alpha_1\,\cos\alpha_1\right)  -\Gamma^2\,\left(\alpha_2-\sin \alpha_2\,\cos\alpha_2\right)  \nonumber \\
 -2\,\mathcal{\tilde{R}}^2\Bigl[\frac{\pi}{2}-\sin\left(\frac{\alpha_1+\alpha_2+2\,\theta_E}{2}\right)\,\cos\left(\frac{\alpha_1+\alpha_2+2\,\theta_E}{2}\right)  \nonumber \\
 -\frac{\alpha_1+\alpha_2+2\,\theta_E}{2}\Bigr]. \nonumber \\
  \label{area}
\end{eqnarray}

\noindent Finally, a last relation between $\alpha_1$ and $\alpha_2$ can be obtained:
\begin{equation}
(\tilde{d}+1)\,\tan\left(\frac{\alpha_1}{2}\right)=(\tilde{d}+\Gamma)\,\tan\left(\frac{\alpha_2}{2}\right). \label{has2}
\end{equation}
Alternatively, this expression can also been rearranged to obtain $\tilde{d}(\Gamma,\alpha_1,\alpha_2)$:
\begin{equation}
\tilde{d}=\frac{\Gamma\,\tan(\alpha_2/2)-\tan(\alpha_1/2)}{\tan(\alpha_1/2)-\tan(\alpha_2/2)}.
\end{equation}

\subsection{Energy balance}

Following Princen's approach, \cite{Princen1970} we neglect the contribution of the radius of curvature in the direction parallel to the fibers. Indeed, far from the menisci, located at both ends of the column, the interface is nearly flat and its contribution is small with respect to the radius of curvature in the cross-section. The variation of energy associated with the addition of an infinitesimal volume ${\rm{d}}\mathcal{V}$ to the column is equal to the work ${\text{d}}W$ done on the system as a result of a pressure difference with the atmosphere:
\begin{equation} \label{energy_1}
{\rm{d}}E={\rm{d}}W=(P-P_0)\text{d}\mathcal{V}=-\frac{\gamma}{\mathcal{R}}\,\mathcal{A}\,{\rm{d}}L,
\end{equation}
where we assume that the increase in volume leads to an increase of length of the column ${\rm{d}}\mathcal{V}= \mathcal{A}{\rm d}L$. In this relation, $\gamma$ is the liquid/vapor surface tension, $P$ and $P_0$ are the pressure values inside and outside the column, respectively. In addition, the increase in length of a column results in the creation of liquid-fiber and liquid-air area while reducing the air-fiber area. This process is associated with an energy variation
\begin{equation}
{\rm{d}}E  =  \left[\left(A_1B_1+A_2B_2\right)\,\left(\gamma_{SL}-\gamma_{SV}\right)+ \left(A_1A_2+B_1B_2 \right)\,\gamma\right]{\rm d}L,
\label{energy_3}
\end{equation}
where $\gamma_{SL}$ and $\gamma_{SV}$ are the solid/liquid and solid/vapor surface tensions, respectively. In this relation $A_1B_1+A_2B_2=2\,\alpha_1 a_1+2\,\alpha_2 a_2$ denotes the region of the fibers initially non-wetted by the liquid and $A_1A_2+B_1B_2=\left(\pi\!-\!2\theta_E\!-\!\alpha_1\!-\!\alpha_2\right)\mathcal{R} $ describes the creation of a liquid-vapor interface. Using the Young's relation, we have $\gamma_{SL}-\gamma_{SV}=\gamma\,\cos\theta_E$. These results lead to an energy change
\begin{equation}
{\rm{d}E}  =  2\Bigl[\left(\pi\!-\!2\theta_E\!-\!\alpha_1\!-\!\alpha_2\right)\mathcal{R}  -2\,(\alpha_1 a_1\!+\!\alpha_2 a_2)\cos\theta_E\Bigr]\gamma{\rm d}L.
\label{energy_2}
\end{equation}
The first term in this expression is proportional to the capillary force at the air-liquid interface and the second term is proportional to the force exerted by the fiber on the liquid. Balancing the expressions (\ref{energy_1}) and (\ref{energy_2}) for the added volume ${\rm d}\mathcal{V} $ and using the expression of the cross-sectional area, we obtain the equation:
\begin{eqnarray}
2\mathcal{\tilde{R}}^2\Bigl[\frac{\pi}{2}-\frac{\alpha_1+\alpha_2+2\,\theta_E}{2} \nonumber \\
+\sin\left( \frac{\alpha_1+\alpha_2+2\,\theta_E}{2} \right)\cos\left( \frac{\alpha_1+\alpha_2+2\,\theta_E}{2} \right)\Bigr] \nonumber \\
+\mathcal{\tilde{R}}\,\Bigl[\left(\sin\alpha_1+\Gamma\,\sin\alpha_2 \right)\,\left(\cos(\alpha_1+\theta_E)+\cos(\alpha_2+\theta_E) \right) \nonumber \\
-2\,\left(\alpha_1+\Gamma\,\alpha_2\right)\Bigr]-\left(\alpha_1-\sin \alpha_1\,\cos\alpha_1 \right) \nonumber \\
-\Gamma^2\,\left(\alpha_2-\sin \alpha_2\,\cos\alpha_2 \right) =0 .  \nonumber \\
\label{eq:ana_solution}
\end{eqnarray}

In the following, we focus on a perfectly wetting fluid, {\textit{i.e.}} with a contact angle $\theta_e=0^{\rm o}$. In this situation, the previous expression becomes
\begin{eqnarray}
2\mathcal{\tilde{R}}^2\left[\frac{\pi}{2}-\frac{\alpha_1+\alpha_2}{2}+\sin\left( \frac{\alpha_1+\alpha_2}{2} \right)\cos\left( \frac{\alpha_1+\alpha_2}{2} \right)\right] \nonumber \\
+\mathcal{\tilde{R}}\,\Bigl[\left(\sin\alpha_1+\Gamma\,\sin\alpha_2 \right)\,\left(\cos\alpha_1+\cos\alpha_2 \right) \nonumber \\
-2\,\left(\alpha_1+\Gamma\,\alpha_2\right)\Bigr]-\left(\alpha_1-\sin \alpha_1\,\cos\alpha_1 \right) \nonumber \\
-\Gamma^2\,\left(\alpha_2-\sin \alpha_2\,\cos\alpha_2 \right) =0 .  \nonumber \\
\label{eq:ana_solution}
\end{eqnarray}


\subsection{Numerical solution}

\begin{figure}[h!]
    \centering
\includegraphics[width=8.5cm]{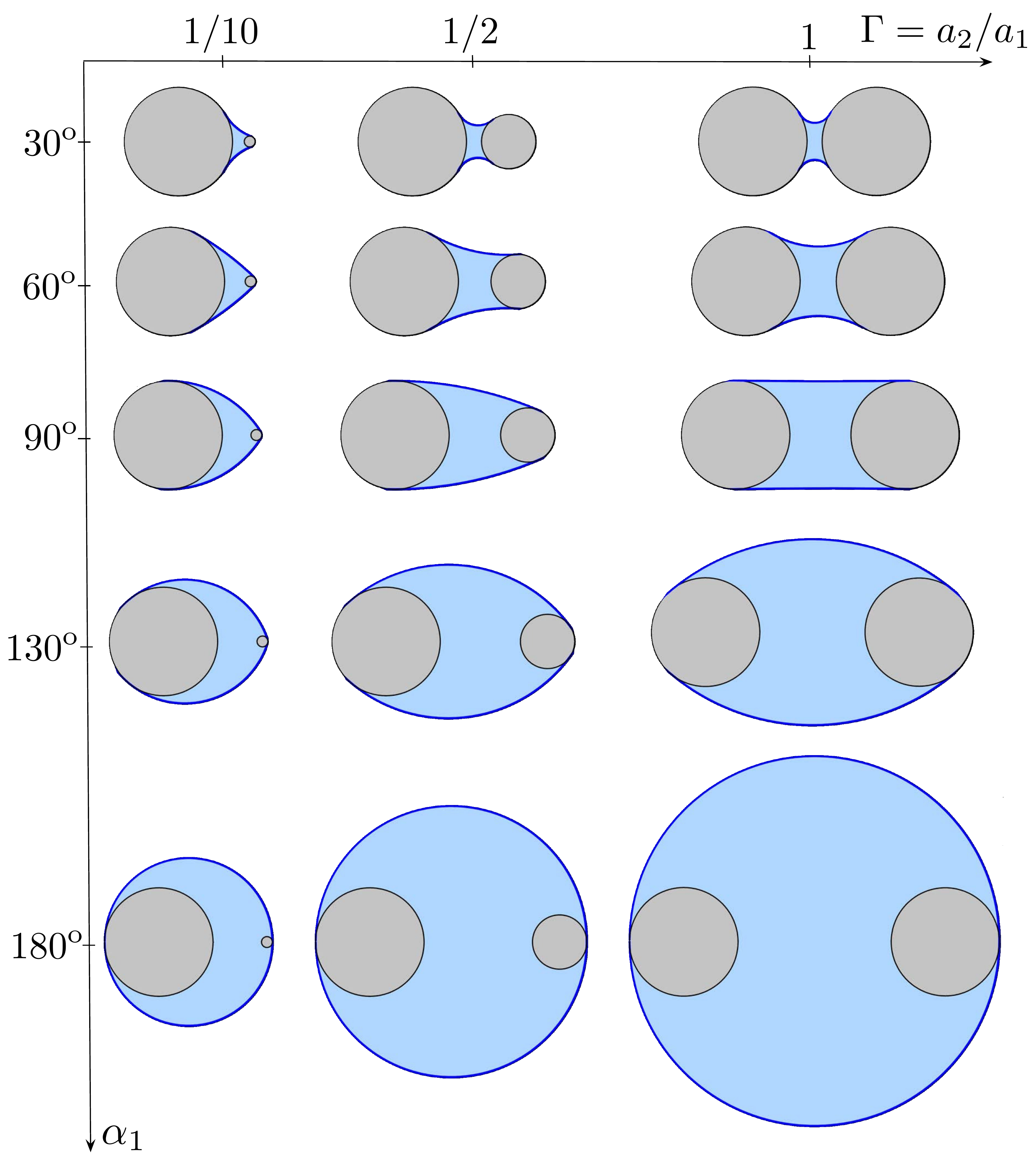}
\caption{Analytical shapes of the cross-section for different radii ratios $\Gamma=a_2/a_1$ and values of the wrapping angle around the largest fiber, $\alpha_1$.}\label{fig:Morphology}
\end{figure}

Considering a perfectly wetting fluid ($\theta_E=0^{\rm o}$), the shape of the cross-section is fully characterized by $\alpha_1$, $\alpha_2$, $\mathcal{\tilde{R}}$ and $\tilde{d}$ for a given value of the fiber aspect ratio $\Gamma=a_2/a_1$. Typically, we want to vary $\alpha_1 \in[0,\,\pi]$ and successively obtain the corresponding values of $\alpha_2$, $\tilde{d}$ and $\mathcal{\tilde{R}}$. The calculations are performed using a custom-written Matlab code\footnote{Available in Supplementary Material (ESI)}. The wrapping angles $\alpha_1$ and $\alpha_2$ are increasing when increasing the inter-fiber distance $\tilde{d}$. When the liquid wraps around the fiber, $\alpha_1$ increases until it reaches a maximum $\alpha_1=\pi$. When $\alpha_1=\pi$, the liquid can not wrap around the fiber anymore therefore leading to a transition from a column morphology to a drop morphology, similarly to the situation of identical parallel fibers.\cite{Princen1970,Protiere2012} In addition, equation (\ref{has2}) shows that for $\alpha_1=\pi$, we have $\alpha_2=\pi$. Substituting these values in equation (\ref{eq:ana_solution}) and (\ref{has22}) we obtain:
\begin{equation}
\tilde{d}_c=\frac{d_c}{a_1}=\sqrt{2\,\Gamma}. \label{dc_analytical}
\end{equation}

Considering two identical fibers leads to $\Gamma=1$ and $\tilde{d}_c=\sqrt{2}$, which recovers the result previously obtained for identical fibers.\cite{Princen1970}

Using the solution of equation (\ref{eq:ana_solution}), we can determine the shape of the cross-section for various radii ratios $\Gamma$ and for increasing values of the wrapping angle $\alpha_1$. In figure \ref{fig:Morphology}, we have represented three situations: $\Gamma=1$, i.e. identical fibers, $\Gamma=1/2$ and $\Gamma=1/10$. We observe that $\alpha_2$ increases faster than $\alpha_1$ for $\Gamma<1$ and that the column can become convex outward for $\alpha_1<90^{\rm o}$. In addition, the shape of the cross-section associated with the maximum area $\mathcal{A}$ is obtained for $\alpha_1=\alpha_2$ for any value of $\Gamma$; only the evolution of the shape with $\alpha_1$ is modified with the radii ratio.

\begin{figure}
    \centering
\includegraphics[width=8.5cm]{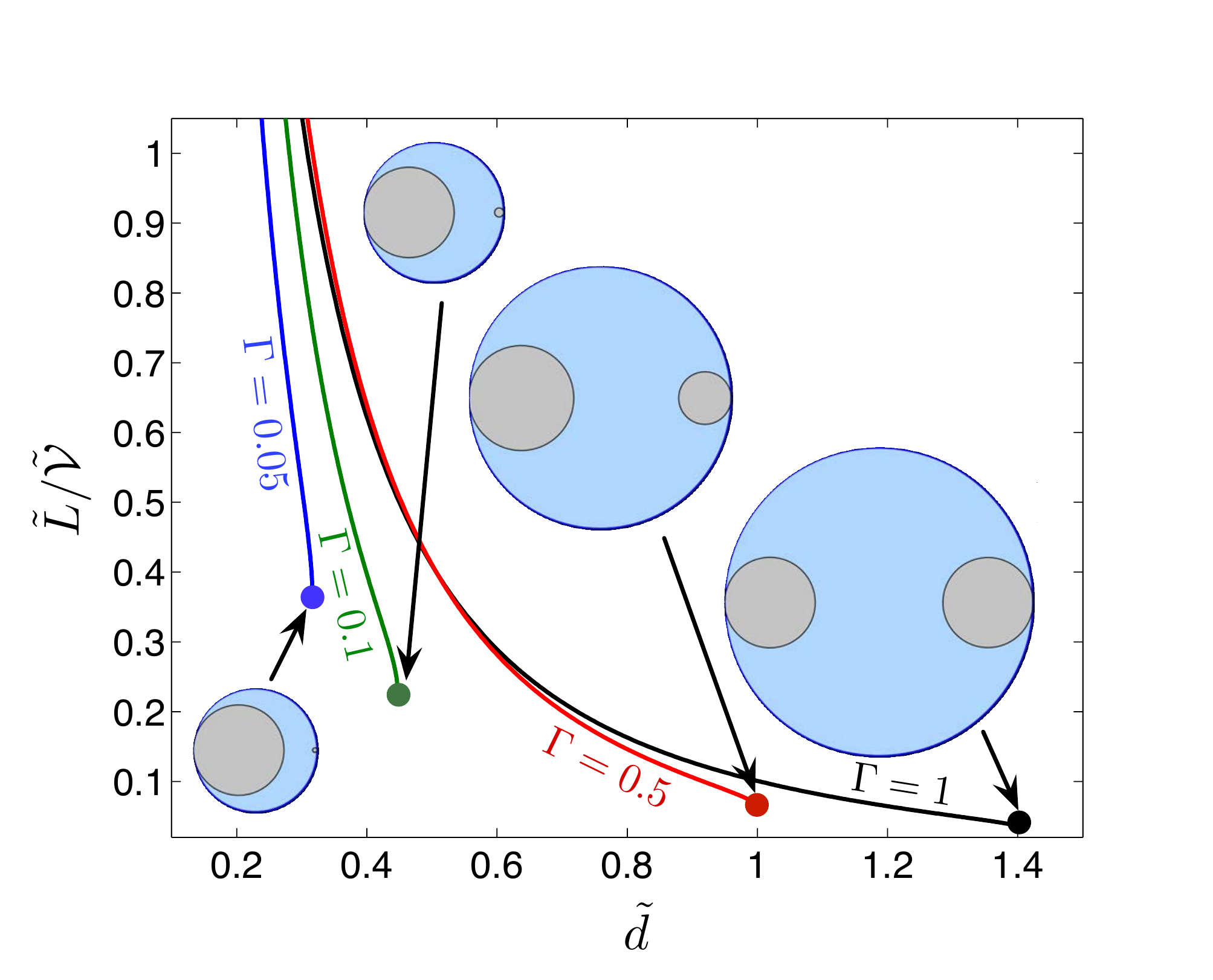}
\caption{Evolution of $\tilde{L}/\tilde{\mathcal{V}}=1/\tilde{\mathcal{A}}$ with the inter-fiber distance $2\,\tilde{d}$ for various fiber aspect ratios $\Gamma=1$ (black) and $\Gamma=0.5$ (red), $\Gamma=0.1$ (green) and $\Gamma=0.05$ (blue). The circles at the end of each curve and the sketches of the cross-sectional area correspond to the maximum inter-fiber distance for the aspect ratio $\Gamma$ considered.}\label{fig:Figure_Theorique_Length}
\end{figure}

For a given value of $\Gamma$ and an inter-fiber distance $\tilde{d}$, we can determine the wrapping angles ($\alpha_1$ and $\alpha_2$) and $\tilde{\mathcal{R}}$, which leads to a unique value of the cross-sectional area $\mathcal{A}$. Consequently, we define the length of a column $\tilde{L}=L/a_1$ for a given volume of liquid $\tilde{\mathcal{V}}=\mathcal{V}/{a_1}^3$ deposited on the fiber. Following Proti\`ere et al.,\cite{Protiere2012} we report the theoretical rescaled parameter $\tilde{L}/\tilde{\mathcal{V}}=1/\tilde{\mathcal{A}}$ in figure \ref{fig:Figure_Theorique_Length}. The influence of the radius ratio $\Gamma$ can be observed both in the value of the critical inter-fiber distance $\tilde{d}_c$ and in the value of $1/\tilde{\mathcal{A}}$. The evolution of $\tilde{L}/\tilde{\mathcal{V}}$ is non trivial: for a fixed $\tilde{d}$, the values of $\alpha_1$ and $\alpha_2$ lead to a smaller cross-sectional area for large radius ratio therefore leading to longer column.

In the next section, we compare our analytical results to experiments performed with different radii ratios $\Gamma$ and a perfectly wetting fluid.


\section{Experiments}

\begin{figure}
    \centering
 \subfigure[]{\includegraphics[width=6.5cm]{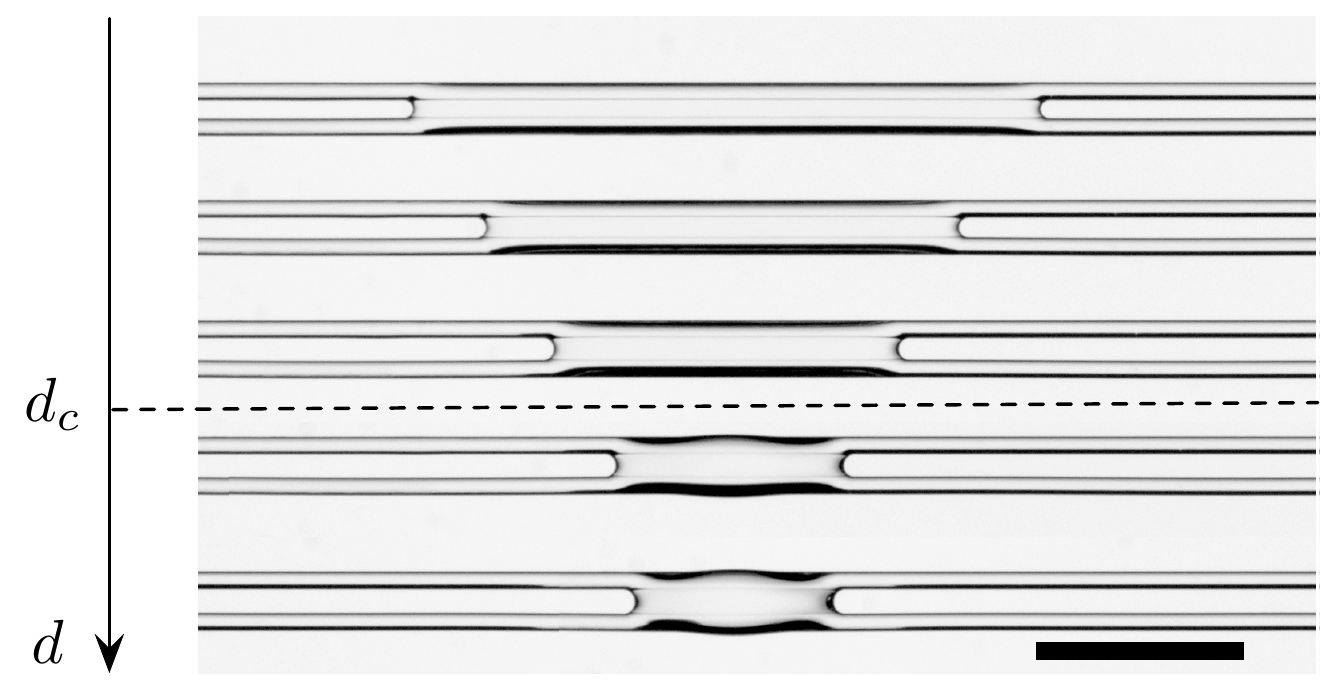}}
  \subfigure[]{\includegraphics[width=6.5cm]{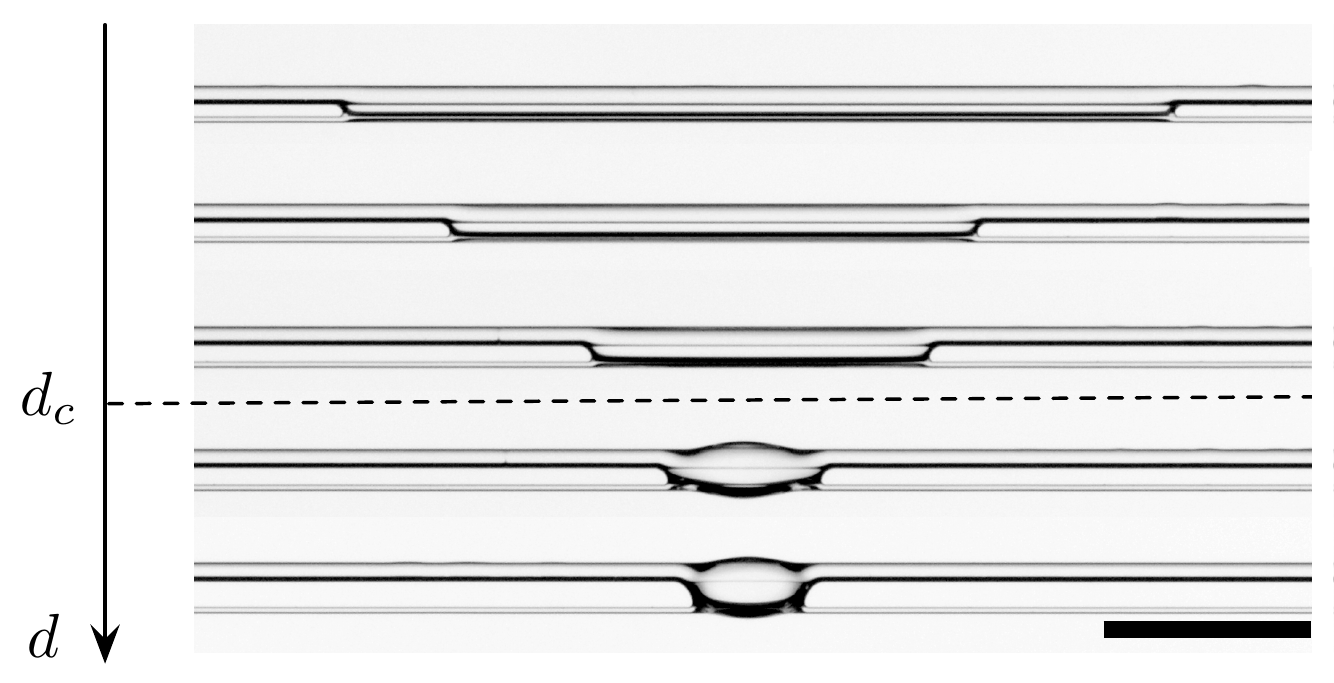}}
  \caption{Evolution of a drop of $5$ cSt silicon oil placed on two parallel nylon fibers as the distance $d$ between the fibers is increased for (a) equal radii $a_1=a_2=225\,\mu{\rm m}$ ($\mathcal{V}=6\,\mu\ell$) and (b) different radii $a_1=225\,\mu{\rm m}$ and $a_2=75\,\mu{\rm m}$ ($\mathcal{V}=3\,\mu\ell$). The dashed line indicates the column to drop transition. Scale bar is 5 cm.}\label{fig:Photo}
\end{figure}

A typical experiment is performed by depositing a small amount of silicone oil (viscosity $5$ cSt), which is a perfectly wetting fluid ($\theta_E=0$), on a pair of nylon fibers of circular cross-section and respective radii $(a_1;a_2) \in [50,225]\,\mu{\rm m}$. Each fiber is clamped on two horizontal micro controllers (PT1, Thorlabs) that allows very accurate increases of $d$ by step of $5\,\mu{ \rm m}$. To minimize the effects of gravity, we deposit a small amount of liquid $\mathcal{V}\in[1,6]\,\mu\ell$ using a micropipette and we neglect the hysteresis induced by the gravitational effects that deform the drop into a hemispherical shape.\cite{Protiere2012} Indeed, a deviation from the analytical prediction is observed when the drop to column transition is reached by reducing the inter-fiber distance $2 \tilde{d}$.\cite{Protiere2012} We therefore start from a small inter-fiber distance $2d$ and increase this distance using the micrometer stage. Pictures are taken from both the top and side views to measure the length of the column and to discriminate the column and drop morphologies. In particular, in the drop morphology, the width of the drop is larger than the spacing of the fibers, which implies that a bead of liquid is located outside the fibers. Through direct measurement, we ensure that the distance between the fibers increases in steps of $10\,\mu{\rm m}$. After increasing the inter-fiber distance, we typically wait for a few minutes so that the equilibrium state is reached prior to imaging.

\begin{figure}
    \centering
 \subfigure[]{\includegraphics[width=8cm]{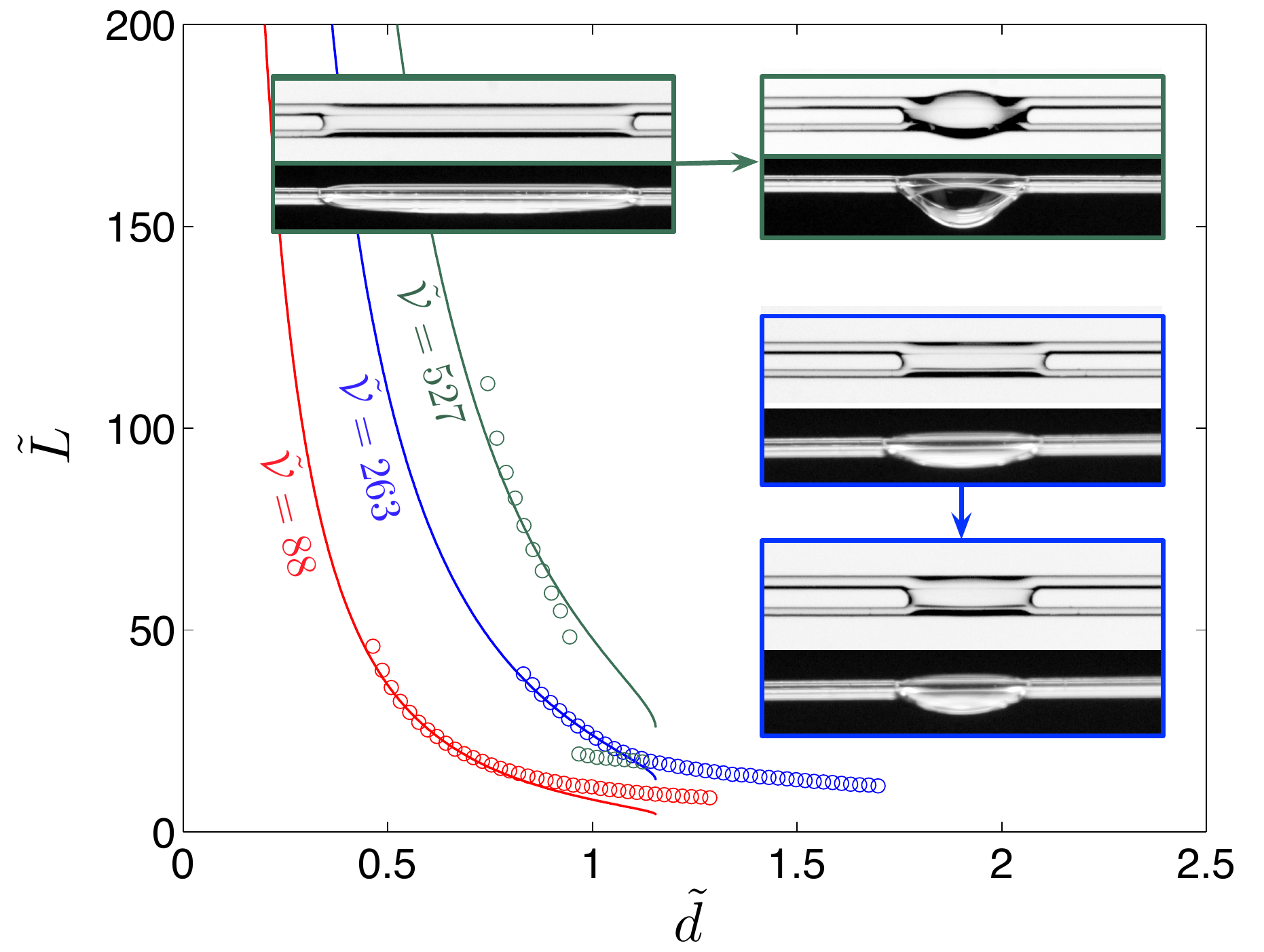}}
 \caption{Evolution of the rescaled length $\tilde{L}=L/a_1$ of $5$ cSt silicon oil placed on two parallel nylon fibers as a function of the rescaled distance $\tilde{d}=d/a_1$ between the fibers for $\Gamma=2/3$ ($a_1=225\,\mu{\rm m}$) and for three different rescaled volumes $\mathcal{\tilde{V}}=88$, $\mathcal{\tilde{V}}=263$ and $\mathcal{\tilde{V}}=527$. The solid line is the prediction of the model described in section \ref{sec:Analytical_Modeling} and the circles are the experimental measurements. The insets show the transition between drop and column for $\mathcal{\tilde{V}}=263$ and $\mathcal{\tilde{V}}=527$ (top pictures: top view; bottom pictures: side view).}\label{fig:theorie_length}
\end{figure}

Typical visualizations for a pair of identical fibers and a pair of different fiber radii are reported in figure \ref{fig:Photo}(a,b), respectively. We observe that the wetting length $\tilde{L}$ decreases in both situations when increasing the inter-fiber distance $2d$. However, the critical distance at which the column to drop transition occurs, $d_c$, is different for the two cases, as predicted by the analytical model. We also observe that the menisci on both extremities of the column become non-symmetric with fibers of different radii (figure \ref{fig:Photo}(b)).

The values of the wetting lengths $L$ measured in both situations are compared to the analytical predictions derived in the previous section. In figure \ref{fig:theorie_length}, we report the evolution of $\tilde{L}$ for various rescaled volumes and non-similar fibers. The analytical model captures the evolution of the length, which further confirms our assumption to neglect the menisci at the ends of the column. For large volumes, the column to drop transition leads to a jump in the measured length, whereas for small liquid volumes this transition remains smooth. Therefore, to estimate the critical distance $d_c$ at which the column to drop transition occurs, we rely on measurements performed with sufficiently large volumes. We should emphasize that, for large volumes, a hysteresis between the column to drop and drop to column transition is observed similar to the report of Proti\`ere et al.\cite{Protiere2012} To predict analytically the value of the hysteresis would require consideration of the energy in the column morphology and in the drop morphology. However the analytical prediction of the drop morphology remains complicated and would necessitate performing numerical simulations that minimize the energy in the presence of gravity using, for instance, Surface Evolver.\cite{Brakke1996,Bedarkar2010,Wu2010,Wu2014}

 \begin{figure}[h!]
    \centering
\includegraphics[width=0.5\textwidth]{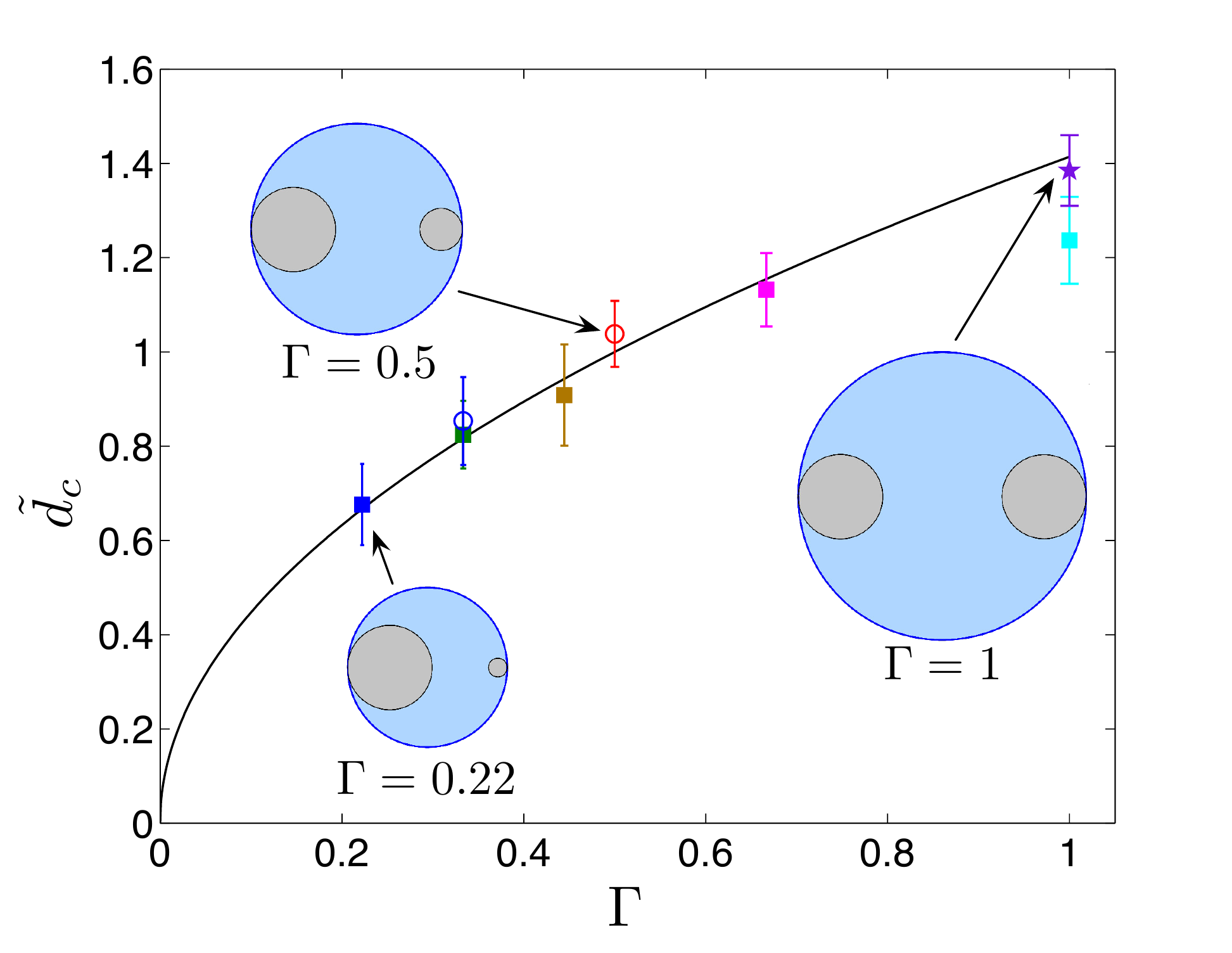}
\caption{Evolution of the critical distance for the column to drop transition when the inter-fiber distance $2d$ is increased for various pair of fibers of radii $a_1$ and $a_2$. The squares are experiments performed with $a_1=225 \,\mu{\rm m}$ and: $a_2=50 \,\mu{\rm m}$ and $V=1 \,\mu{\rm \ell}$ (blue), $a_2=75 \,\mu{\rm m}$ and $V=1 \,\mu{\rm \ell}$ (green), $a_2=100 \,\mu{\rm m}$ and $V=3 \,\mu{\rm \ell}$ (brown), $a_2=150 \,\mu{\rm m}$ and $V=3 \,\mu{\rm \ell}$ (magenta), $a_2=225 \,\mu{\rm m}$ and $V=6 \,\mu{\rm \ell}$ (cyan). Hollow circles shows experiments with with $a_1=150 \,\mu{\rm m}$ and: $a_2=50 \,\mu{\rm m}$ and $V=1 \,\mu{\rm \ell}$ (blue), $a_2=75 \,\mu{\rm m}$ and $V=1 \,\mu{\rm \ell}$ (red). The purple star is data obtained from Proti\`ere et al. for $a_1=a_2=125 \,\mu{\rm m}$ and $V=4 \,\mu{\rm \ell}$ (blue).\cite{Protiere2012} The continuous line is the analytical prediction (\ref{dc_analytical}).}\label{fig:figure_d_c}
\end{figure}

Starting from a column shape, we can measure the critical inter-fiber distance $2d_c$ at which the column to drop transition occurs and compare these experimental results to our analytical prediction. Analytically, we find that for a perfectly wetting fluid the transition occurs for $\tilde{d}_c=\sqrt{2\,\Gamma}$ (equation \ref{dc_analytical}). We have performed experiments with  pairs of fibers $a_1$ and $a_2$ including the situation described by Princen,\cite{Princen1970} $a_1=a_2$ leading to $\Gamma=1$. A summary of our experimental results is presented in figure \ref{fig:figure_d_c}. We observe that for all values of $\Gamma$ the analytical prediction captures well the morphological transition, which further confirms our analytical modeling.


\section{Discussion and conclusion}

In this paper, we have presented an analytical model, confirmed by our experiments, of the column morphology and of the critical inter-fiber distance for the column to drop transition. These results extend a previous analytical study devoted to identical fibers\cite{Princen1970} to the situation of fibers of different radii.

Although the present paper focuses on the situation of parallel fibers, since it constitutes a simple case that has been well studied in the literature,\cite{Princen1970,Protiere2012} the analytical calculation is straightforward to extend to the situation of crossed-fibers. Indeed, for crossed fibers, the main difference is that the inter-fiber distance $2d$ depends on the distance $y$ to the ``crossing'' point:\cite{Sauret2014}
\begin{equation}\label{toto1}
\tilde{d}(y)=\frac{d(y)}{a_1}=\sqrt{\frac{y^2}{{a_1}^2}\,\tan^2\left(\frac{\delta}{2}\right)+\frac{(1+\Gamma)^2}{4}}-\frac{(1+\Gamma)}{2},
\end{equation}
where $\delta$ is the tilt angle between the fibers. It follows that the area of the cross-section depends on the varying inter-fiber distance $2d(y)$ according to the relation (\ref{area}). Therefore, we can determine $\mathcal{A}$ for all values of $\tilde{d} \leq \sqrt{2\,\Gamma}$. The condition $\tilde{d}_c = \sqrt{2\,\Gamma}$ predicts the maximum length of the column $L_c$. The length of the column for a given volume of fluid $\tilde{\mathcal{V}}$ deposited on the crossed-fibers satisfies: \begin{equation}
\mathcal{\tilde{V}}  =  \int_{-L}^{L}\,{\mathcal{\tilde{A}}}(y)\,\text{d}y.
\end{equation}
If $L > L_c$ the fluid is either in a mixed morphology or in a drop shape whereas for $L < L_c$ the fluid is in a column morphology. With this geometric picture that characterizes the morphology, we could also determine the liquid-air area and thus the evolution of the drying rate for the different morphologies.\cite{Duprat2013,Boulogne2015}

Similarly, the calculation could be extended easily to fibers of identical radius but with different surface properties, i.e. if the contact angle of the fluid is different for the two fibers $\theta_{E,1} \neq \theta_{E,2}$. Similar behavior to the results presented in this paper is expected to happen: namely, the wrapping angle on one fiber, $\alpha_1$, is different from the wrapping angle on the second fiber $\alpha_2$. It will modify the transition between the drop and column morphologies as well as the critical inter-distance $\tilde{d}_c$ compared to fibers of identical radius and the same surface properties.

To conclude, we have taken a step toward the description of wetting in array of fibers by considering the liquid morphology for fibers of different sizes. Work is in progress to better describe the distribution of liquid in a large network of fibers. First, as emphasized above, numerical simulations, for instance using Surface Evolver \cite{Brakke1996}, are needed to predict the shape of a drop with and without gravity and determine the associated energy. Finally, more complex interactions, i.e., between three or more fibers, should be considered.

\newpage
\appendix

\section{Appendix A: cross-sectional area of the liquid column}

\begin{figure}[h!]
    \centering
\includegraphics[width=0.35\textwidth]{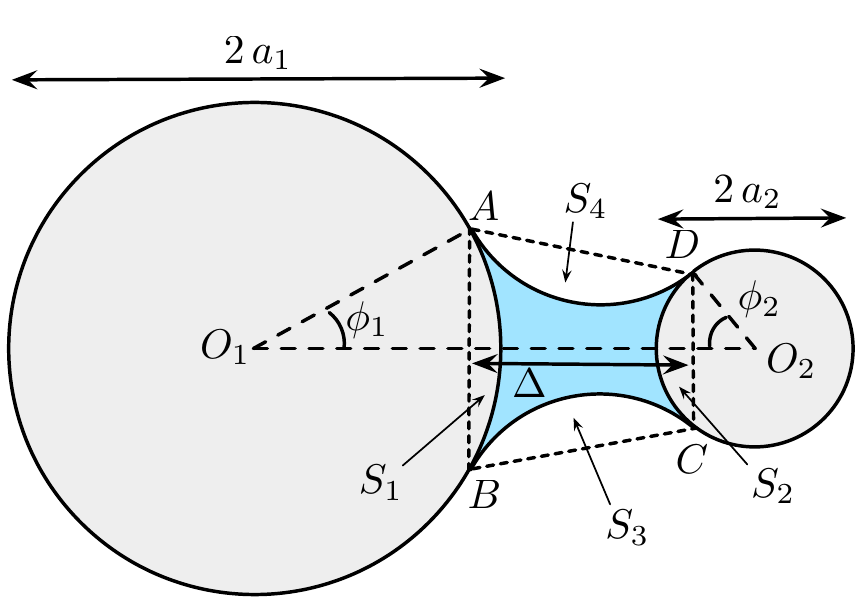}
\caption{Schematic of the cross-section to determine its area.}\label{fig:aire}
\end{figure}

To determine the area of a cross section ABCD, we consider the area of the trapezoid shown in Fig. \ref{fig:aire}:
\begin{equation}
\mathcal{A}_1=\frac{(AB+CD)\,\Delta}{2}
\end{equation}
with $AB=2\,a_1\,\sin\alpha_1$, $CD=2\,a_2\,\sin\alpha_2$ and $\Delta=2d+a_1\,(1-\cos\alpha_1)+a_2\,(1-\cos\alpha_2)$. We then remove the areas S1, S2, S3 and S4, which are given respectively by:

\begin{eqnarray}
S_1& =& {a_1}^2\,\left[\alpha_1-\sin\alpha_1\,\cos\alpha_1\right] \\
S_2 & = & {a_2}^2\,\left[\alpha_2-\sin\alpha_2\,\cos\alpha_2\right] \\
S_3 & = &\mathcal{R}^2\,\Bigl[\frac{\pi}{2}-\frac{\alpha_1+\alpha_2+2\,\theta_E}{2} \nonumber \\
& & -\sin\left(\frac{\alpha_1+\alpha_2+2\,\theta_E}{2}\right)\,\cos\left(\frac{\alpha_1+\alpha_2+2\,\theta_E}{2}\right)\Bigr] \\
S_4 & =& S_3.
\end{eqnarray}

Therefore, we obtain the cross-sectional area
\begin{eqnarray}\label{appen_A}
\mathcal{A}  = \left( a_1\,\sin\alpha_1+a_2\,\sin\alpha_2\right)\,\left[a_1\,(1-\cos\alpha_1)+a_2\,(1-\cos\alpha_2)+2d\right] \nonumber \\
-{a_1}^2\,\left(\alpha_1-\sin \alpha_1\,\cos\alpha_1\right)  -{a_2}^2\,\left(\alpha_2-\sin \alpha_2\,\cos\alpha_2\right)  \nonumber \\
 -2\,\mathcal{R}^2\Bigl[\frac{\pi}{2}-\sin\left(\frac{\alpha_1+\alpha_2+2\,\theta_E}{2}\right)\,\cos\left(\frac{\alpha_1+\alpha_2+2\,\theta_E}{2}\right)  \nonumber \\
 -\frac{\alpha_1+\alpha_2+2\,\theta_E}{2}\Bigr]. \nonumber \\
\end{eqnarray}
In addition, relation (\ref{has22}) can be rewritten
\begin{equation}
{(1-\cos\alpha_1)+\Gamma(1-\cos \alpha_2)+2\tilde{d}} =\mathcal{\tilde{R}}[{\cos(\alpha_1+\theta_E)+\cos(\alpha_2+\theta_E)}].
\end{equation}
Substituting this expression in relation (\ref{appen_A}) leads to
\begin{eqnarray}\label{appen_A}
\frac{\mathcal{A}}{{a_1}^2}  = \mathcal{\tilde{R}}[{\cos(\alpha_1+\theta_E)+\cos(\alpha_2+\theta_E)}]\left(\sin\alpha_1+\Gamma\,\sin\alpha_2\right) \nonumber \\
-\left(\alpha_1-\sin \alpha_1\,\cos\alpha_1\right)  -\Gamma^2\,\left(\alpha_2-\sin \alpha_2\,\cos\alpha_2\right)  \nonumber \\
 -2\,\mathcal{\tilde{R}}^2\Bigl[\frac{\pi}{2}-\sin\left(\frac{\alpha_1+\alpha_2+2\,\theta_E}{2}\right)\,\cos\left(\frac{\alpha_1+\alpha_2+2\,\theta_E}{2}\right)  \nonumber \\
 -\frac{\alpha_1+\alpha_2+2\,\theta_E}{2}\Bigr].
\end{eqnarray}

\noindent Considering two identical fibers of radii $a_1=a_2=a$ leads to $\alpha_1=\alpha_2$ and the cross-sectional area simplifies to
\begin{eqnarray}
\frac{\mathcal{A}}{a^2}=4\,\mathcal{\tilde{R}}\,\sin\alpha\,\cos(\alpha+\theta_E)-2\,\left(\alpha-\sin \alpha\,\cos\alpha\right) \nonumber \\
-2\,\mathcal{\tilde{R}}^2\,\left[\frac{\pi}{2}-\alpha-\theta_E-\sin\left(\alpha+\theta_E\right)\,\cos\left(\alpha+\theta_E\right)\right],
\end{eqnarray}
which is the expression established by Princen for a pair of identical fibers.\cite{Princen1970}

\section*{Acknowledgements}

FB acknowledges that the research leading to these results received funding from the People Programme (Marie Curie Actions) of the European Union's Seventh Framework Programme (FP7/2007-2013) under REA grant agreement 623541. HAS thanks the Princeton MRSEC for partial support of this research. ED is supported by set-up funds by NYU Polytechnic School of Engineering.

\newpage
    \bibliography{article}
    \bibliographystyle{unsrt}

    \end{document}